\documentclass[conference]{IEEEtran}
\usepackage{cite}
\usepackage{multirow}

\usepackage[linesnumbered,ruled,vlined]{algorithm2e}
\usepackage{hyperref}
\usepackage{multirow}
\usepackage{booktabs}
\usepackage{todonotes}
\usepackage{amsmath,latexsym,amssymb,graphicx,dsfont,amsthm,amsfonts}

\usepackage{comment}

\begin{document}
\title{PAN-DOMAIN: Privacy-preserving Sharing and Auditing of Infection Identifier Matching }

\author{\IEEEauthorblockN{Will Abramson, William J, Buchanan, Sarwar Sayeed, Nikolaos Pitropakis,  Owen Lo\IEEEauthorrefmark{1}, 
}
\IEEEauthorblockA{\IEEEauthorrefmark{1}Blockpass ID Lab, School of Computing, Edinburgh Napier University, Edinburgh}
}

\maketitle

\thispagestyle{plain}
\pagestyle{plain}

\begin{abstract}
The spread of COVID-19 has highlighted the need for a robust contact tracing infrastructure that enables infected individuals
to have their contacts traced, and
followed up with a test. The key entities involved within a contact tracing infrastructure may include the Citizen, a Testing Centre (TC), a Health Authority (HA), and a Government Authority (GA). Typically, these different domains need to communicate with each other about an individual.
A common approach is when a citizen discloses his
personally identifiable information to both the HA a TC, if the test result comes positive, the information is used by the TC to alert the HA. Along with this, there can be other trusted entities that have other key elements of data related to the citizen. 
However, the existing approaches comprise severe flaws in terms of privacy and security. 
Additionally, the aforementioned approaches are not transparent and often being questioned for the efficacy of the implementations. 
In order to overcome the challenges, this paper outlines the PAN-DOMAIN infrastructure that allows for citizen identifiers to be matched amongst the TA, the HA and the GA. PAN-DOMAIN ensures that the citizen can keep control of the mapping between the trusted entities using a trusted converter, and has access to an audit log. 

\end{abstract}

\IEEEpeerreviewmaketitle

\section{Introduction}



Contact tracing applications have been under development in many countries throughout the world~\cite{world_contact_tracing}, however; the efficacy and ethical implications of these solutions are being questioned all the time~\cite{parker2020ethics}. In particular, the ability to automatically alert individuals that they have come into contact with an infected individual through an application. Leading cryptography and privacy experts have spoken out about this, stating that the response needs to be led by epidemiologists and public health experts rather than technologists~\cite{schneier_tracing, real_world_tracing}. While at the same time, privacy concerns of centralised solutions that open up new avenues for surveillance to both state and non-state actors are well-recognised~\cite{eff_privacy_concerns,taylor2020data}. There needs to be an appropriate balance between centralisation and separation of concerns between different domains in order to protect individual privacy. Additionally, any tracing system needs to be transparent, accountable and driven by human decisions rather than code decisions. 

Digital contact tracing involves the creation of a unique identifier, which can be used to generate a set of rolling identifiers that get advertised over Bluetooth beacons. Infected daily identifiers are then published, which can be used to generate the rolling identifiers and check to see if they match any stored in the application - a potentially infectious contact. Apple and Google thus worked together to create an API for Bluetooth to track physical proximity between phones. 
This method supports the creation of a daily diagnosis key, which links to a rolling identifier that can be generated from the daily diagnosis key~\cite{Wired_Amazon_Apple}. However, this comes with some serious flaws. Some concerns do raise questions, such as what is to stop someone from reporting to be infectious when indeed they are not? What about those who are infected but decide not to notify the tracing application? There needs to be a balance between user control, the validity of data, transparency and accountability.

PAN-DOMAIN puts forward a novel solution for the linking of identifiers within pandemic contact tracing~\cite{camenisch2015linkable, camenisch2017privacy}, which integrates into the Google and Apple Bluetooth tracing method~\cite{Apple} but enforces that only valid tests from a testing centre can cause a contact to be traced digitally. Unlinkable pseudonyms allow a citizen to be known by many unique, unlinkable pseudonyms across all the health and government services they interact with. The system implements a central converter capable of blindly converting between a citizen's identifier from two different datasets. It also has an efficient, privacy-preserving mechanism for the user to be able to audit which linking services have been requesting pseudonym conversions with other services. This paper contributes to  
outlining a concrete and realistic example of this cryptography applied to pandemic tracing.

The main contributions of this paper can be summarised, as follows:
 
\begin{itemize}
\item It briefly analyses the existing solutions to determine their flaws, thus revealing the causes of their weaknesses. 
\item It introduces a novel architecture that links identities with the domains ensuring that citizens confidential data remain secured.
\item On top of the architecture, it integrates the Google/Apple privacy-preserving method to construct citizen's identifiers.
\end{itemize}

\section{Related work}\label{Sec:related}

\begin{figure}
 \centering
  \includegraphics[width=0.3\textwidth]{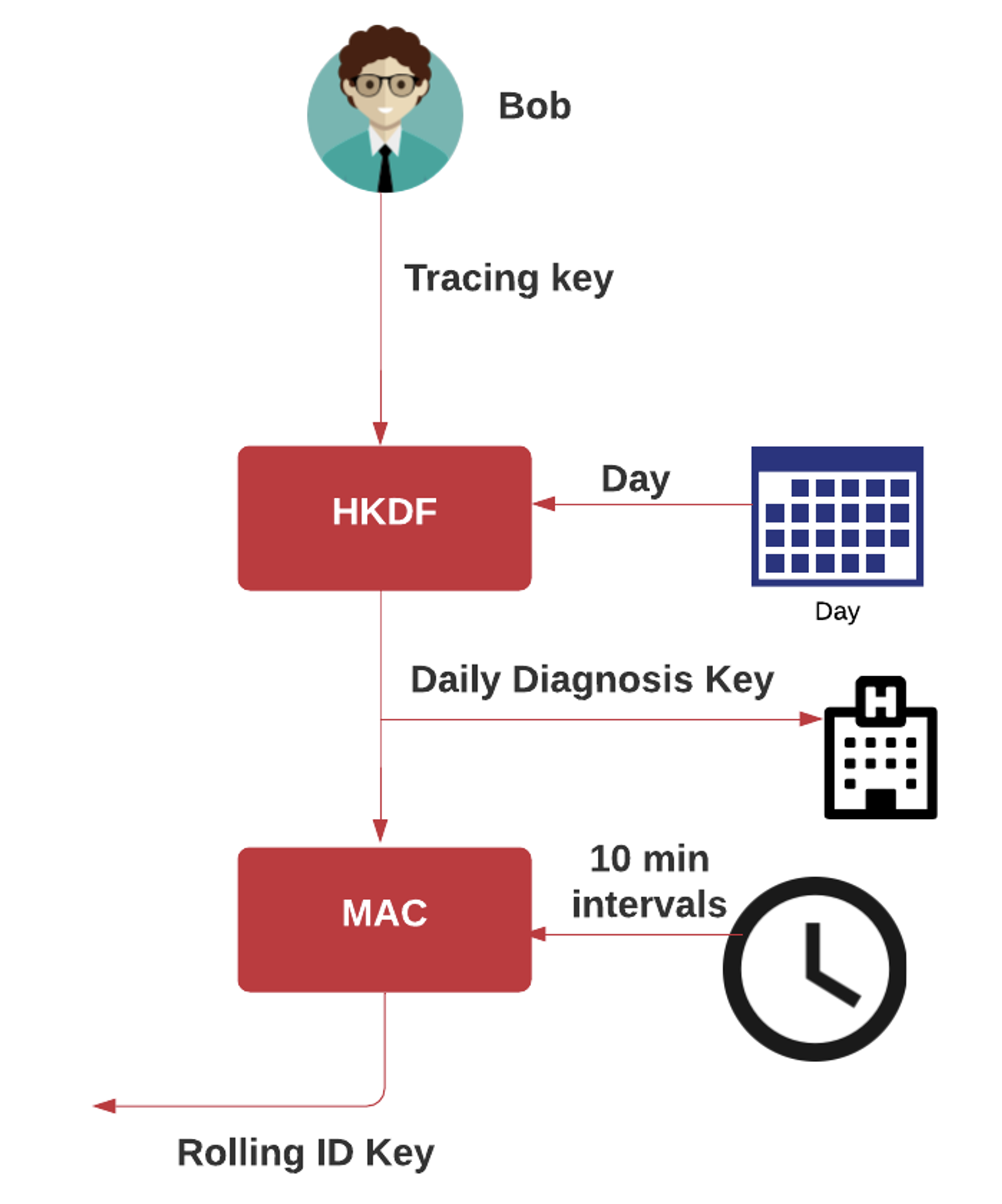}
 \caption{Google/Apple API}\label{fig:apple}
 \vspace{-0.8cm}
 \end{figure}

Figure~\ref{fig:apple} illustrates an overview of the Apple/Google API~\cite{AppleGoogle2020}. 
The method works by Bob generating a unique 256-bit tracing key for his phone and where this key must be kept secret. Bob creates a daily tracing key (diagnosis key), by creating a hash from the tracing key and the current day. From this hash, it should not be possible to determine his tracing key (as it is generated from the random 256-bit tracing key). Every 10 minutes, he then creates a rolling ID key, and which is an HMAC identifier (a signed hash) of his daily tracing key and a counter for the number of 10 minutes that have passed that day.

The Pan-European Privacy-Preserving Proximity Tracing Initiative (PEPP-PT) proposes an open-source Bluetooth based platform sharing software, standards and services that can be utilised for the development of COVID-19 contact tracing Apps. Each national health authority can tweak the software according to its own policies and processes. The software aims at measuring proximity data and alerting the traced contacts of a user if detected positive to COVID-19 while adhering to privacy. As an EU initiative, it has a wide approach across national borders~\cite{PePP}.

When maintaining multiple sets of personal data, being able to correlate an individual across these data sets is often viewed as a requirement. The common solution to this challenge is to provide a global identifier to an individual and then to use this as a primary key across all data stores, thus making the correlation of an individual by any entity is trivial. However, such a solution has serious flaws, with the first being that
any entity can correlate any individual in any other data set. This comes from the lack of mechanisms for preventing or recording such a correlation. Secondly, the use of a single identifier in many data sets leaves the individual vulnerable to a security breach in any of these data sets. Once the global identifier has been revealed, an attacker would be capable of correlating the identifier in all other datasets.


Control and authorisation of data requests are necessary for many settings. The Belgium system deals with this by implementing a centralised hub for all data exchange whose role is to authorise all data exchanges across services~\cite{belgium_app}. This has serious privacy concerns; no one central authority learns all requests made about a user.

When someone is being tested for COVID-19, the process should reveal as little information about that person as possible. Testing is often given over to private contractors who the public are less likely to place trust in them having their personal data — the reason we have to give over our personal data when testing is for the purpose of identification. When a test comes back positive, there is a need to notify the individual tested. By utilising CL-17~\cite{camenisch2017privacy} it is possible for the responsibility testing and notification of results to be separated such that testing facilities no longer need to request personal information, but instead request an identifier generated through a specific converter.

Typically, the governments create a data lake, where all of the data is linked and merged. This might relate to health records, tax records, and so on. Hence, this would allow the health care provider to see individuals tax return, and the tax inspector may be able to see the health record causing serious threats to individuals confidential information. 
An improved solution is thus for citizens' identifiers across domains to be unlinkable except with the express permission of the citizen. Linking occurs in a zero-knowledge fashion such that even after linking neither domain knows the identifier of that user in the other domain. Two of the best methods proposed for this matching are known as CL-15~\cite{camenisch2015linkable} and CL-17~\cite{camenisch2017privacy}. Within these papers, Camenisch and Lehmann outline a way that the user can control their own ID (UID), and then store a pseudonym on System A (PsIDA), and another one on System B (PsIDB). A converter then links the two IDs together, but cannot identify the person involved.


The system described in CL-15~\cite{camenisch2015linkable} and improved in CL-17~\cite{camenisch2017privacy} enables an individual to generate a unique, unlinkable identifier, often called a pseudonym, for a particular server that needs an identifier to recognise and remember the individual. This pseudonym is generated through a process involving a third party, which Camenisch and Lehmann call \emph{a converter}. 
The CL-17 paper added a crucial transparency aspect to the converter, enabling it to publish an encrypted log of each conversion request to a public bulletin board such that only the individual with whom the request relates to can decrypt and view the log. This gives individuals the capability to audit which servers requested conversions; such capabilities are now required by the recent GDPR legislation. This solution enables audit capabilities without compromising individual privacy is an achievement. It is one of the main reasons we believe unlinkable pseudonym can add value to many different domains, healthcare being a prime example. 

Our work differs from all existing approaches because it comprises a method that authorises a citizen to interact with other domains leveraging a unique  unlinkable pseudonyms. Hence, the citizens will be within a trusted mechanism where the converter is able to convert between the identifier and datasets. This allows a citizen to relate to other domains without disclosing any
confidential information. It is a way of securing citizen's data and decline any unwanted requests they may receive. For instance, the Testing Centres (TC) can be an untrusted private entity who will only be given access to an identifier generated through an honest-but-curious ID converter in a blind fashion. 
Thus, the proposed system provides an optimal balance between the competing architectures that is often described in contact tracing scenarios.

\section{Contact Tracing Overview}\label{Sec:contract_overview}

\begin{figure*}
 \centering
  \includegraphics[width=0.60\textwidth]{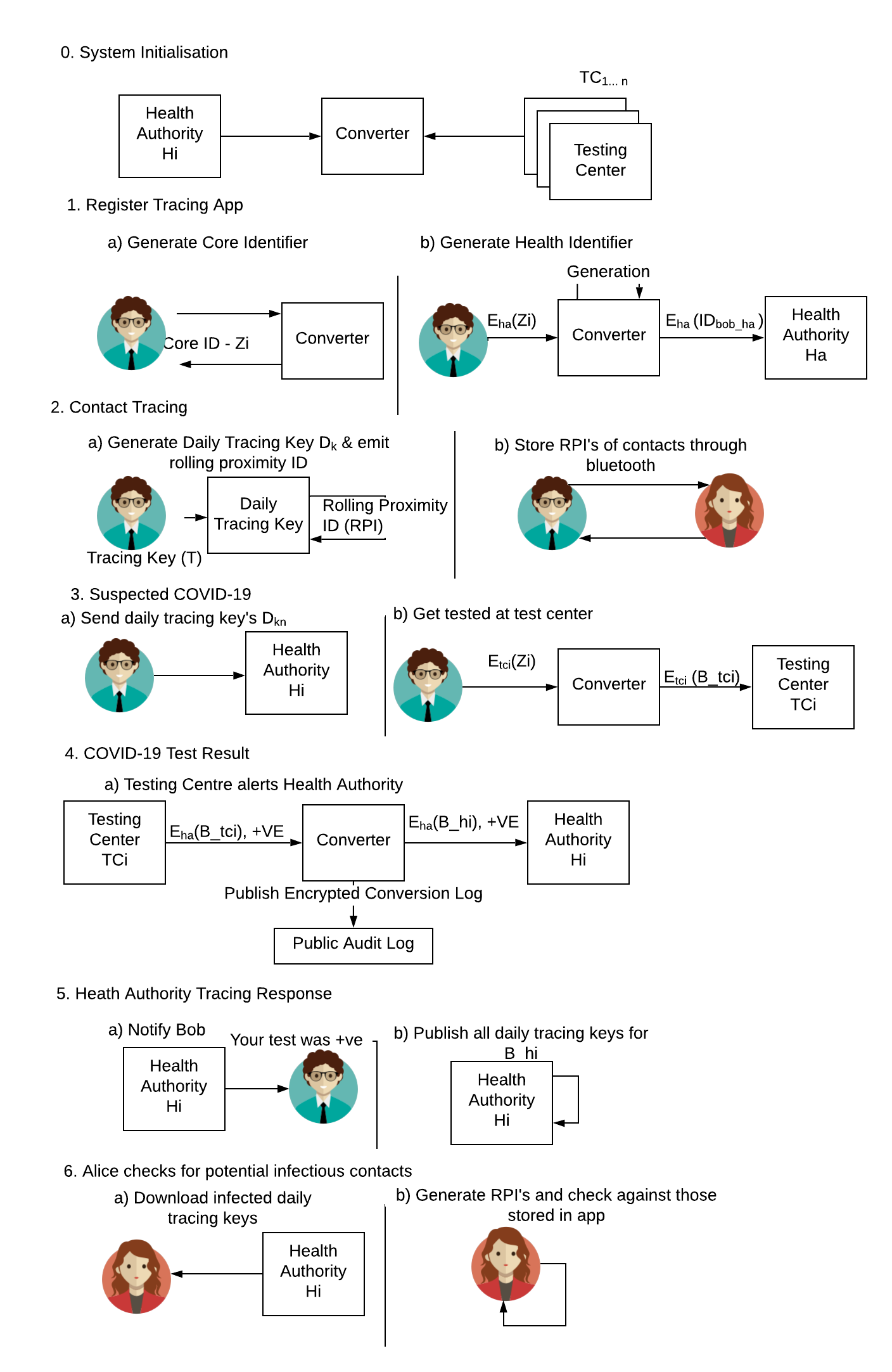}
 \caption{CL-17 and Contact Tracing Scenario}\label{fig:overview}
  \vspace{-0.8cm}
 \end{figure*}

In the scenario considered, a health authority is responsible for developing a mobile tracing application and broadly follows the Apple/Google method described in Section~\ref{Sec:related}. 
The overview of the different stages of this protocol is depicted in Figure~\ref{fig:overview}.

\subsection{Initial Setup}

Before the Health Authority can deploy a tracing application combined with the CL-17 linking functionality, the system must go through some initial configuration where the ID converter and each of the servers much generate some secret information. The converter generates common information for the system, including the list of servers, the testing centres and health authority in this case, and the domain the pseudonyms should be generated within. This is distributed to all participants of the system. For each server, the converter then generates a secret exponent used to create server-specific identifiers from a users core identifier, and a signing key pair is used to create signatures specific to that server on each identifier or conversion the converter issues. This signature prevents the converter from converting identifiers that it did not issue. Each server also needs to generate a set of keys and secret information. They generate a key for a pseudo-random permutation; this permutation is applied to all unencrypted identifiers it receives from the converter adding an additional layer of randomness. They also create an encryption key pair and a signing key pair.

\subsection{Download and Register for Tracing Application}

\begin{figure*}
        \center{\includegraphics[width=1\textwidth]
        {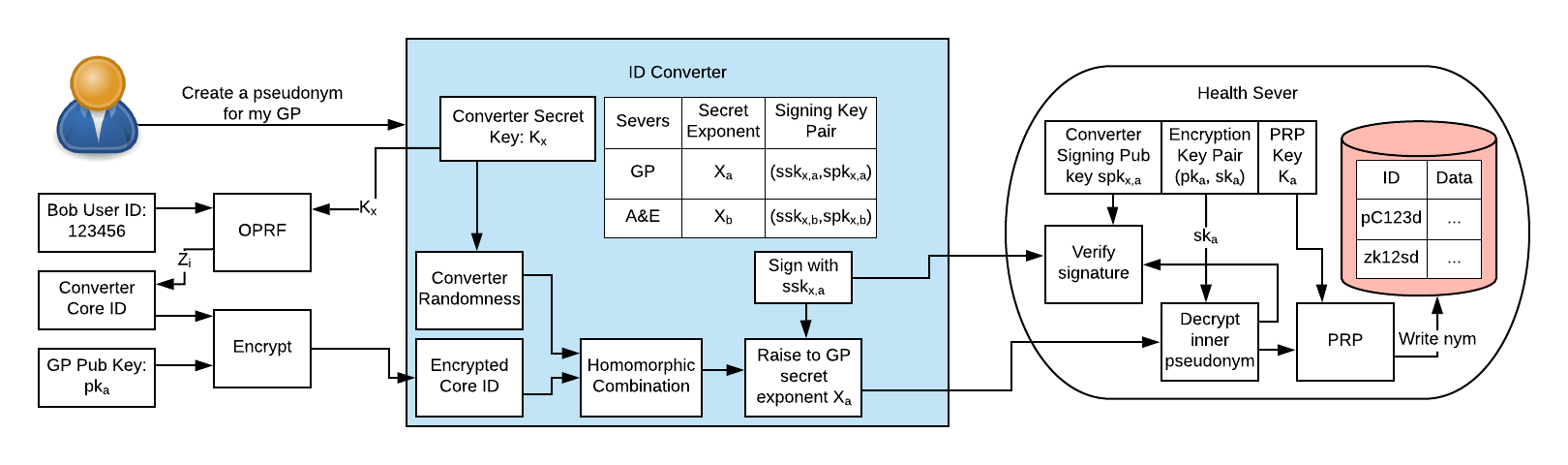}}
    \caption{Overview of generation process}
    \label{fig:gen}
  \vspace{-0.8cm}
\end{figure*}

The tracing application is a mobile app which is already downloaded and runs on Bob's phones. This process involves registration with the Health Authority, rather than divulge PII such as name and email, a pseudonym is generated in an interactive manner. First, Bob creates a core identifier of $Z_i$ to be used within this domain. This is created with the ID Converter through an oblivious pseudorandom function. The converter never learns the identifier of the user it interacts with. Next, Bob uses $Z_i$ to generate a pseudonym for the Health Authority in a blind fashion. Only the Health Authority learns this identifier and they use it to correlate all subsequently related tracing data for Bob. Additionally, Bob's application generates a tracing key $k$ for use in the contact tracing algorithm. Figure~\ref{fig:gen} provides an overview of how a pseudonym is generated blindly by a converter, for a more detailed mathematical version see CL-17.

{}

\subsection{Contact Tracing}\label{sec:bobalice}

Bob's tracing application generates a Daily Tracing Key $D_k$ from his tracing key $K$. The Daily Tracing Key is used to generate new Rolling Proximity Identifiers (RPI's) every ten minutes. These RPI's are emitted over a Bluetooth beacon to all devices in the vicinity listening. So when Bob meets Alice for a walk, if both individuals are using the tracing App, then they share and store each other's RPI's. These can be used later to verify proximity to a potentially infectious contact.

\subsection{Bob suspects he has a case of COVID-19}

The day after meeting Alice, Bob starts to develop symptoms the could be associated with a case of COVID-19. He uses his tracing application to notify the Health Authority and provide them with the Daily Tracing Keys of the past few days and immediately visits a drive-through testing centre. Using the tracing application, specifically Bob's core identifier $Z_i$, Bob generates an identifier through the converter for this testing centre. The testing centre correlates Bob's test to this identifier and await results. Bob returns home and self-isolates awaiting the results of his test.

The Health Authority and the tracing application may also decide to publish Bob's daily tracing keys in this situation. They could do this with the clear indication that the status of these keys is not yet known, but that they could potentially represent an infectious contact. Rapid processing of test results would greatly increase confidence and effectiveness of such a situation.

\subsection{Bob's COVID-19 Test Result}

\begin{figure*}
 \centering
  \includegraphics[width=0.9\textwidth]{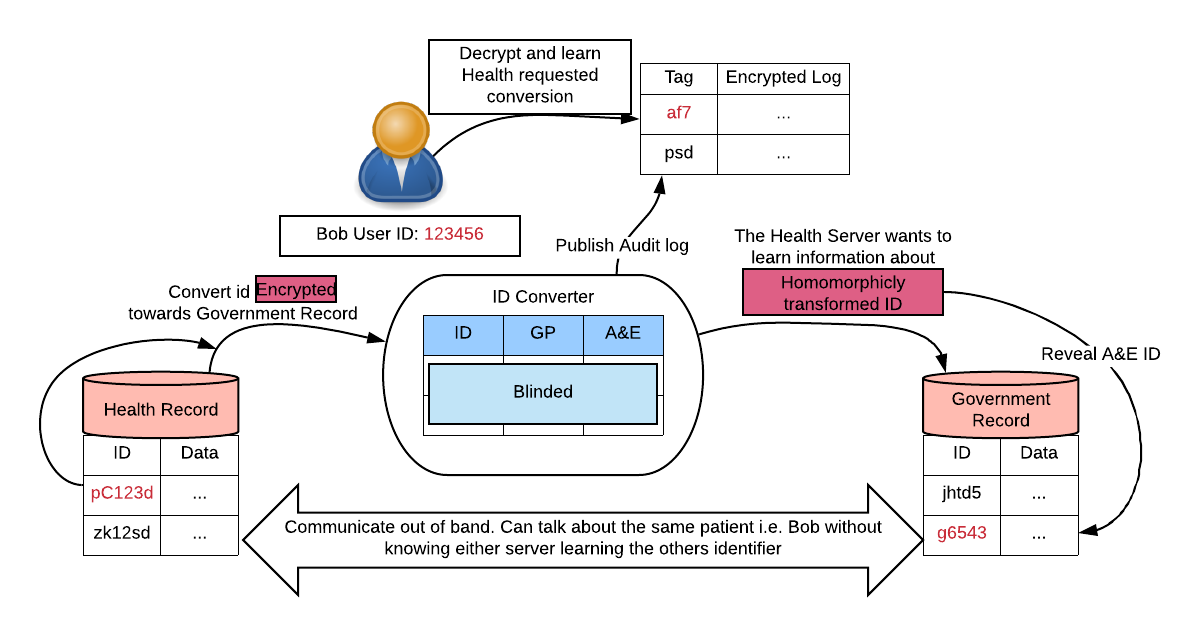}
 \caption{Pseudonym Conversion (High Level)}
 \label{fig:highlevelconversion}
  \vspace{-0.8cm}
 \end{figure*}

The Testing Center gets the results back from Bob's COVID-19 test; however, no confidential information about Bob is accessible.  They just know that some identifier, $123avbd3$, and has received a positive test result. This identifier was generated by Bob through the ID Converter during the testing procedure. This means that the testing centre requests a pseudonym conversion into the domain of the Health Authority. On the input of a correctly structured conversion request, the converter converts from a nym under one server to a nym in another server. It does this blindly using homomorphic encryption based on El Gamal. Once both of the servers know the identifier that communication is referring to the Testing Center can notify the Health Authority, the results of the individual. The power of CL-17 ensures that both the Testing Center and the Health Authority can communicate about the same individual without needing to share the same unique identifier and without the converter learning anything other than a Test Center wants to convert into the Health Authority domain. Pseudonym Conversion is depicted in Figure \ref{fig:highlevelconversion}.
.

\subsection{Health Authority Tracing Response}

The Health Authority is now aware that an individual identified by a certain identifier has just tested positive according to some a Testing Center. The Health Authority using this identifier to first notify the user, Bob, through the tracing App that they have tested positive and should self isolate. Then assuming Bob has already started providing the Health Authority with Daily Tracing Keys, these are then published as known infected keys. If Bob has not been sending these keys,  they can be requested from him through the App specifying the recommended number of daily keys previous to testing that should be provided. Note, the App cannot force Bob to divulge this information, however, it is expected that those who download such an app would be willing to use it in a way specified by Health Authorities assuming the trust has been developed in its use over time.

\subsection{Alice Check's for Infectious Contact}

Through the contact tracing application, Alice downloads all of the newly published Daily Tracing Key's by the Health Authority. She then uses each of the keys to generate the full set of RPI's. The tracing App compares the generated RPIs to the ones the App has saved through proximal contact with individuals she has met, including Bob in Section~\ref{sec:bobalice}. Due to her meeting with Bob, a number of RPIs from one of Bob's daily keys that got published by the Health Authority matches the RPIs stored on Alice's phone. Alice is notified that she may have had an infectious contact and should isolate immediately and seek a test. Alice is able to complete the steps in Section \ref{sec:bobalice} quickly verifying if she is infected or not and tracing her contacts as necessary.


The above scenario provides a high-level overview of how contact tracing can be done in an authenticated and privacy-respecting manner. Tracing only occurs when a test result has been verifiable positive as attested by a known Testing Center. 
%
%
It also shows how this model could be further extended to include a citizen's primary healthcare provider. 


\section{Auditing Protocol and Integration}\label{Sec:audit_protocol}

\begin{figure*}
 \centering
  \includegraphics[width=0.65\textwidth]{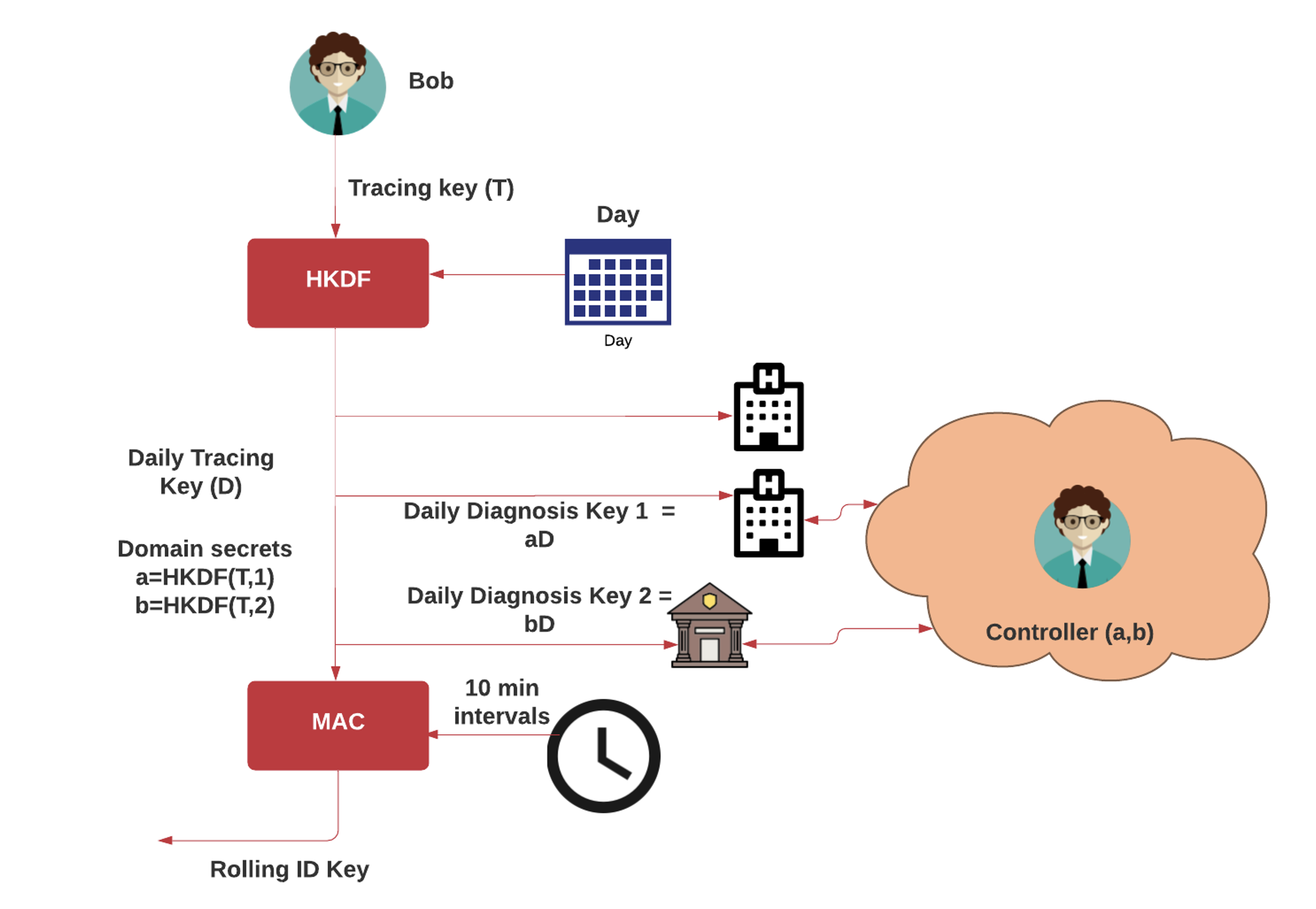}
 \caption{PAN-DOMAIN Overview}\label{fig:pan}
  \vspace{-0.8cm}
 \end{figure*}

\subsection{Auditing}\label{sec:conversion}

\begin{figure*}[!htb]
        \center{\includegraphics[width=1\textwidth]
        {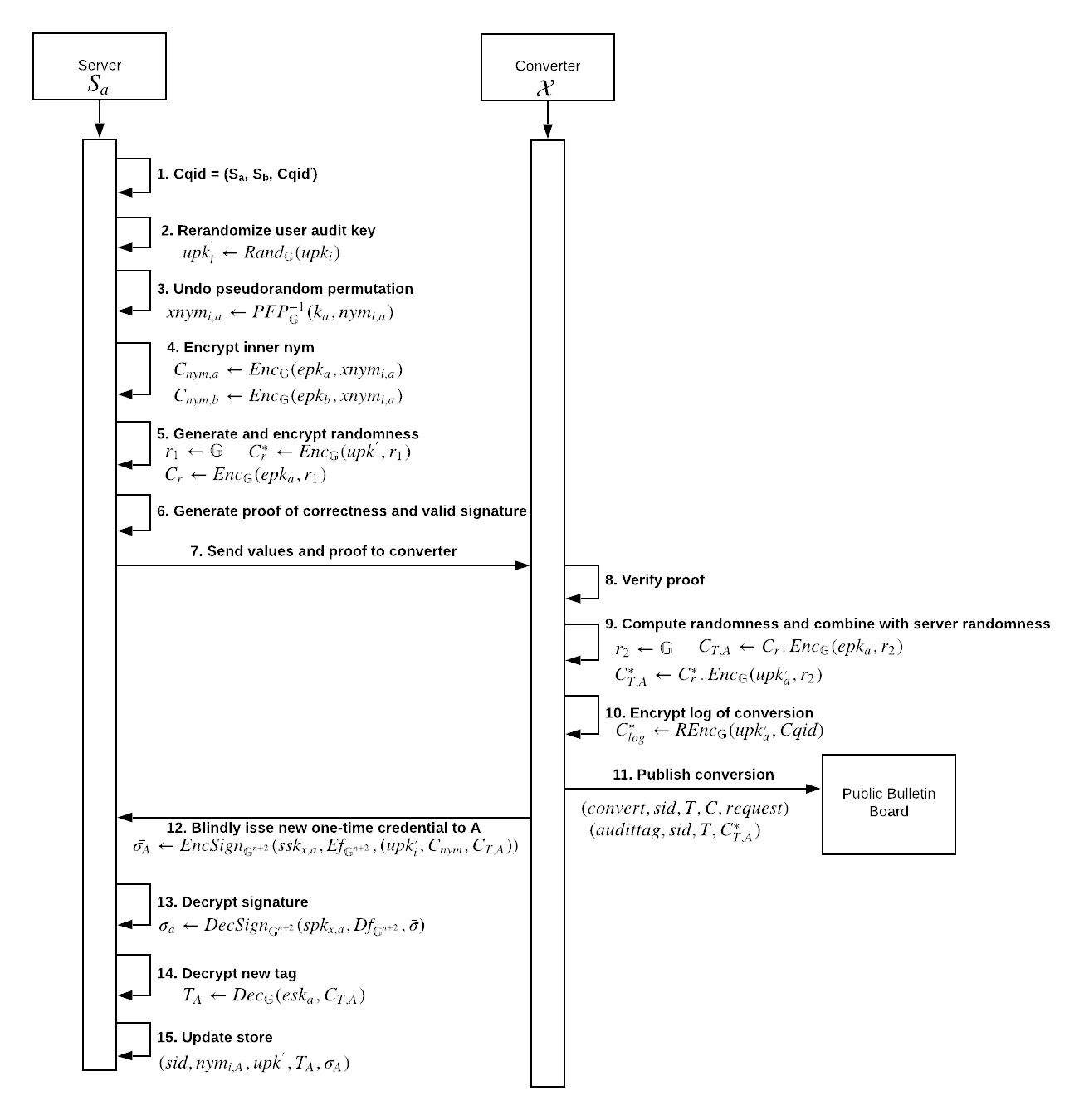}}
    \caption{Auditing protocol}
    \label{fig:audit}
     \vspace{-0.8cm}
\end{figure*}

Figure~\ref{fig:audit} provides an overview of PAN-DOMAIN architecture. While the pseudonym mapping method works well for privacy, we still have a problem, and that is related to transparency. With this Bob wants to know who is snooping on him, and for the system to log that someone in the HMRC has linked his health record. In Estonia and Japan (“MyPortal”)\cite{kawai2004my} this is a core method used to increase trust, as all accesses to government records from officials are logged and the user can then prompt as to why the person accessed their records.

In the European Data Protection Directive, too, one of the three principles includes the right for a data subject to be informed when their personal data is being processed. Thus, transparency is key to creating an infrastructure where trust can be built. Within a traditional linker system, this would be fairly easy, as Alice makes a request to the linker service (Trent), who then links Bob’s identities between the data infrastructure. At the same time, Trent will log Alice’s access and the reason for it. Bob can then see the reason that she access the linking of the data, and query if he thinks that it is unacceptable.

\begin{figure}
\center{\includegraphics[width=0.5\textwidth]
        {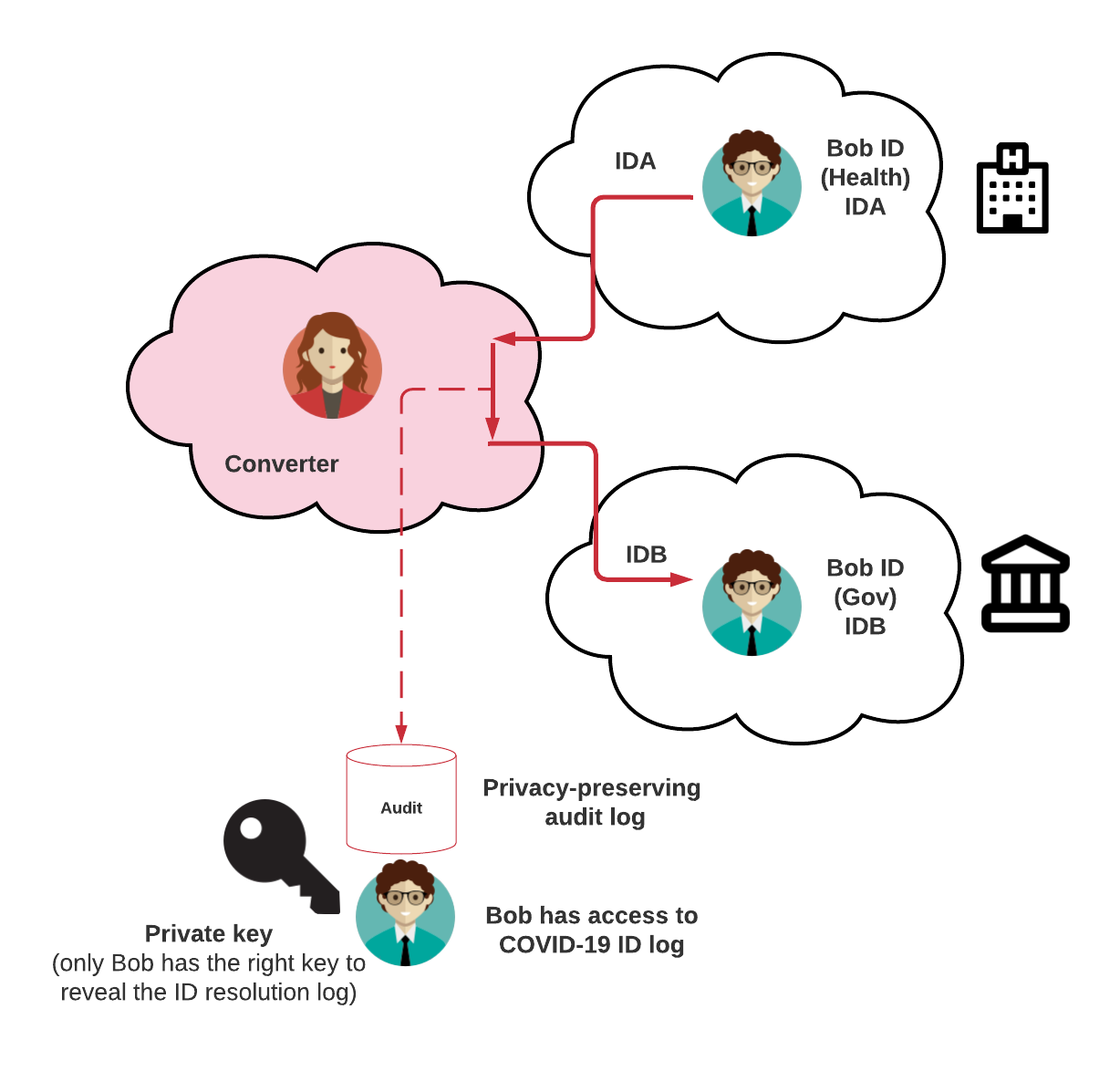}}
    \caption{Privacy-preserving audit log}
    \label{fig:au02}
 \vspace{-0.8cm}
\end{figure}

This method can be seen as having a weak point, as it leaves behind a trail of accesses which could be used to breach Bob’s privacy. Camenisch~\cite{camenisch2017privacy} thus proposes using the blind linker method~\cite{camenisch2015linkable} and then create a trusted audit service and which logs the blind mappings (Figure \ref{fig:au02}). In this way, the converter saves to an audit log system but makes sure that all the pseudonyms are unlinkable. Bob is then the only person who can reveal the linkages, and trace that Alice has been matching his records. The method in \cite{camenisch2017privacy} works by creating a public key for each pseudonym, and where Bob has the associated private key to reveal these. 

{}

\subsection{Method}

A method is required which to integrate with a COVID-19 contact tracer, and which integrates with the methods defined in the previous sections. Overall the CL methods were developed using discrete logarithms, and are often not efficient given prime numbers of over 2,048 bits. A more efficient solution is thus to use elliptic curve methods. One of the best implementations is to use Curve 25519, and which uses a base point at $x=9$ and a prime number of $2^{255}-19$. Initially, we start with a random 256-bit number value for Bob and store on his mobile phone:

\begin{equation}
ID_{Bob}={0,1}^{256}  
\end{equation}

As shown in Figure~\ref{fig:pan}, we first use the Google/Apple method of anonymising Bob's ID \cite{Apple}, and creating a Daily Tracing Key (DTK) and then a Rolling Proximity ID (RPI). Initially Bob creates a random ID value on his phone ($ID_Bob$). Then when he is diagnosed with COVID-19, he generates a DTK using his ID, a counter for the day and an HKDF (HMAC Key Derivation Function) \cite{eronen2010hmac}:

\begin{equation}
\textit{Diag}_{Bob}=\textit{HKDF}(ID_{Bob},Day)   
\end{equation}

This creates a daily diagnosis key which, when passed to a health authority (HA), allow them to trace Bob's device. Every 10 minutes, an RPI is then created using a 10 minute counter ($MinIntervals$) and the DTK and which creates a MAC (Message Authentication Code):

\begin{equation}
\textit{RPI}_{Bob}=\textit{MAC}(Diag_{Bob},MinsIntervals)   
\end{equation}

If the HA has the $Diag_{Bob}$ key they can match this to the $RPI_{Bob}$ that is changed every 10 minutes. This model assumes only one domain (such as the NHS in the UK), but let's say we have two or more interest parties, and we don't want them to have the same DTK for Bob. For this we can now modify the method so that we have two DTKs:

\begin{equation}
\textit{Diag}_{Bob}=\textit{HKDF}(\textit{ID}_{Bob},Day) 
\end{equation}

We now create two identities for Bob using two random values ($a$ and $b$). The identity given to the first domain is:

\begin{equation}
\textit{Diag}_{Bob1}=a \times \textit{Diag}_{Bob} \times G  
\end{equation}

and to the second domain:

\begin{equation}
\textit{Diag}_{Bob2}=b \times \textit{Diag}_{Bob} \times G  
\end{equation}

and where $G$ is the base point on Curve 25119, and $a$ and $b$ are scalar values. So, if we have two domains of the NHS and the Police, each of these would be given one of the diagnosis keys. Now, we create a matching service between the two, and which will link the identities. In the NHS receives $Diag_{Bob1}$, and there needs to be a link into the Police domain, the matching service much translate $Diag_{Bob1}$ into $Diag_{Bob2}$. First, the NHS and the Police will perform a key exchange using ECDH (Elliptic Curve Diffie Hellman) in order to generate a secret:

\begin{equation}
secret=\textit{ECDH}(NHS_{keypair},Police_{keypair}) 
\end{equation}

NHS, using Curve 25519, defines Bob's DTK with:

\begin{equation}
\textit{NHS}_{Bob}= (a \times \textit{Diag}_{Bob} \times G)
\end{equation}

It will now encrypt this with the secret value, and pass to the Proxy:

\begin{equation}
\textit{Proxy}_{Bob}=secret \times (a \times \textit{Diag}_{Bob} \times G)
\end{equation}

and where $secret$ is the random scalar value created in the key exchange, and $G$ is the base point on Curve 25519. Next, the Proxy multiplies this value by $b$ value and multiplies this by the inverse of $a$. This is a homomorphic multiplication way to produce:

\begin{equation}
\textit{EncPolice}_{Bob}=secret \times (b \times \textit{Diag}_{Bob} \times G)
\end{equation}

The Police domain then decrypts this by multiplying by the inverse of the secret value to give:

\begin{equation}
\textit{Police}_{Bob}=b \times \textit{Diag}_{Bob} \times G
\end{equation}

In this way, we now have a privacy-preserving link between the two domains \cite{Buchanan}.

\subsection{Encryption Method Conversion}

Within the original CL-15 paper \cite{camenisch2015linkable}, the authors used the partial homomorphic property of RSA with discrete logarithms (ElGamal) to convert the identities. If we have $xa$ and $xb$, and we use RSA, the encrypted value becomes:

\begin{equation}
E(IDA) = {ID^{xa}}^e \pmod N = ID^{xa e} \pmod N    
\end{equation}

If we then multiply by xb/xa, we then perform a homomorphic conversion to:

\begin{equation}
E(IDB) = {ID^{xb}}^e \pmod N 
\end{equation}

We can then just use the RSA private key to decrypt. The usage of discrete logarithms can have performance issues, thus Table \ref{tab01} outlines the results of 1,000 conversions for different homomphonic encryption methods. With Curve 25519, we use multiplicative homomorphic encryption. Full homomorphic encryption (FHE) allows for a range of homomophic operations, including multiplication and division. In this case, we evaluation HEAAN (Homomorphic Encryption for Arithmetic of Approximate Numbers) \cite{cheon2017homomorphic}.  The hardware used is AMD Phenom II X6 10055T, and compiled using Golang. It can be seen from the results that 512-bit and 1,024-bit ElGamal conversion work better than elliptic curve methods, but for 2,048 bits and above, there is a significant overhead. In fact, the lattice method for HEAAN actually works more efficiently than 4,096-bit ElGamal conversion.

\begin{center}
\begin{table}
\centering
\begin{tabular}{c c}
 \hline
 Method &   Average time (for 1,000 conv)\\
 \hline\hline
 Curve 25519 &   599.03 ms\\
  \hline
 HEAAN & 4.04s\\
  \hline
 ElGamal (512-bit) & 338.2177ms \\
  \hline
 ElGamal (1,024 bit) & 508.93ms\\
  \hline
 ElGamal (2,048 bit) & 2.25s\\ 
  \hline
 ElGamal (4,096 bit) & 6.93s\\ 
 \hline
 
 \hline
\end{tabular}
\vspace{0.5em}
\caption{Evaluation of ID conversions}\label{tab01}

\end{table}
\end{center}


\section{Conclusions}\label{Sec:conclusions}
The COVID-19 pandemic has proved that there is a need for a sturdy contact tracing system which can be used to identify an individual infected person, and then trace the people that this infected person has been in contact with. While this paper has focused on a technical architecture to protect the privacy of individuals identified and represented within such a system using a novel cryptographic protocol, the authors want to emphasise that such a technical system could only be successful is the social infrastructure of contact tracing is effective and trusted. Unfortunately countries managing their public health in response to COVID-19 have repeatedly turned to technologies as \textbf{the} solution, in what has been described as \textit{technology theatre} that can serve to papers over the cracks in society but often ends up amplifying existing power dynamics that created them in the first place \cite{taylor2020data}.

\bibliographystyle{IEEEtran}
\bibliography{bibliography}

\end{document}